\documentstyle[aps]{revtex}

\newcommand{\bea}{\begin{eqnarray}}
\newcommand{\eea}{\end{eqnarray}}
\newcommand{\n}{\noindent}

\begin{document}

%%%%%%%%%%%%%%%%%%%%%%%%%%%%%%%%%%%%%%%%%%%%%%%%%%%%%%%%%%%%%%%%%
\draft

\title{Newtonian Limits of the Relativistic Cosmological Perturbations}
\author{Jai-chan Hwang}
\address{Department of Astronomy and Atmospheric Sciences,
         Kyungpook National University, Taegu, Korea}
\author{Hyerim Noh}
\address{Korea Astronomy Observatory,
         San 36-1, Whaam-dong, Yusung-gu, Daejon, Korea}
\date{\today}
\maketitle

%%%%%%%%%%%%%%%%%%%%%%%%%%%%%%%%%%%%%%%%%%%%%%%%%%%%%%%%%%%%%%%%%
\begin{abstract}

Relativistic cosmological perturbation analyses can be made
based on several different fundamental gauge conditions.
In the pressureless limit the variables in certain gauge conditions
show the correct Newtonian behaviors.
We consider the general curvature and the cosmological constant in the
background medium.
The perturbed density in the comoving gauge, and the perturbed velocity 
and the perturbed potential in the zero-shear gauge show the same behavior 
as the Newtonian ones in a general scale.
Far inside horizon, except for the uniform-density gauge, density perturbations
in all the fundamental gauge conditions show the correct Newtonian behavior.
In this paper we elaborate these Newtonian correspondences.
We also present the relativistic results considering general pressures
in the background and perturbation.

\end{abstract}

%%%%%%%%%%%%%%%%%%%%%%%%%%%%%%%%%%%%%%%%%%%%%%%%%%%%%%%%%%%%%%%%%
\section{Introduction} 
                                    \label{sec:Introduction}

The analysis of gravitational instability in the expanding universe
model was first presented by Lifshitz (1946) in a general relativistic context. 
Historically, the much simpler, and in hindsight, more intuitive 
Newtonian study followed later (Bonner 1957).
The pioneering study by Lifshitz is based on a gauge choice which is
commonly called as the synchronous gauge. 
As the later studies have shown, the synchronous gauge is only 
one way of fixing the gauge freedom out of several available
fundamental gauge conditions 
(Bardeen 1980; Kodama $\&$ Sasaki 1984; Hwang 1991b).
As will be summarized in the following, out of the several gauge conditions
only the synchronous gauge fails to fix the gauge mode completely, thus often 
needs more involved algebra.
Though, as long as one is careful of the algebra
this gauge choice does not cause any kind of intrinsic problem.
The gauge condition which turns out to be especially suitable for
handling the perturbed density is the comoving gauge (Nariai 1969; Sakai 1969).
Since the comoving gauge condition completely fixes the gauge transformation
property, the variables in this gauge can be equivalently considered as
gauge invariant ones.
As mentioned, there exist several such fundamental gauge conditions
each of which completely fixes the gauge transformation properties.
Thus, the variables in such gauge conditions are equivalently gauge invariant.
Using the gauge freedom as an advantage for handling the problem
was emphasized in Bardeen (1988).
In order to use the gauge choice as an advantage a {\it gauge ready method}
was proposed in Hwang (1991b) which will be adopted in the following.

The variables which characterize the self gravitating Newtonian fluid flow are
the density, the velocity and the gravitational potential 
(the pressure is specified by an equation of state).
Whereas, the relativistic flow may be characterized by various components
of the metric (or curvature) and the energy momentum tensor.
Since the relativistic gravity theory is a constrained system we have the 
freedom of imposing certain conditions on the metric or the energy 
momentum tensor as coordinate conditions.  
In the perturbation analyses the freedom arises because we need to introduce
a fictitious background system in order to describe the physical perturbed
system.
The correspondence of a given spacetime point between the perturbed spacetime
and the ficticious background one could be made with certain degrees 
of freedom.
This freedom can be fixed by the suitable gauge conditions 
based on the spacetime coordinate transformation.
Studies in the literature show that a certain variable in a certain gauge
condition correctly reproduces the corresponding Newtonian behavior.
Although the perturbed density in the comoving gauge shows the Newtonian
behavior, the perturbed potential and the perturbed velocity in the
same gauge do not behave like the Newtonian ones; e.g., in the comoving
gauge the perturbed velocity vanishes by the coordinate (gauge) condition.
It is known that both the perturbed potential and the perturbed velocity
in the zero-shear gauge correctly behave like the corresponding
Newtonian variables (Bardeen 1980).

In this paper we will elaborate establishing such correspondences 
between the relativistic and Newtonian perturbed variables. 
Our previous work on this subject is presented in Hwang (1994a; H1 hereafter)
and Hwang $\&$ Hyun (1994).
In the following, we will derive the relativistic equations which describe 
the perturbed density, potential and velocity variables in several
available gauge conditions and will compare the equations with the
corresponding equations satisfied by the Newtonian system.
In our analyses we will include both the spatial curvature and 
the cosmological constant in the background medium.
Based on such correspondences we will extend our result to the situations
with general pressures in the background and perturbations.
We will present the relativistic equations satisfied by the gauge invariant
combinations and will derive the general solutions valid in the large
scale considering both the spatial curvature ($K$) and the 
cosmological constant ($\Lambda$).

In \S \ref{sec:Newtonian} we present the closed form equations
and {\it general solutions} for the Newtonian perturbed variables in 
a prssureless medium.
The solutions are valid for general $K$ and $\Lambda$.
In \S \ref{sec:Relativistic} we summarize a complete set of
equations describing the relativistic perturbations with general
pressures (including both the adiabatic, entropic and anisotropic ones).
The equations are presented in a gauge ready form and 
the method of handling the gauge issue is briefly described.
In \S \ref{sec:Six-gauges} we consider a pressureless limit.
We derive the equations for the density, the potential and the 
velocity in several different fundamental gauge conditions.
By comparing these relativistic equations in various gauges with
the Newtonian ones in \S \ref{sec:Newtonian}, 
in \S \ref{sec:Correspondences} we identify the gauge conditions
which reproduce the correct Newtonian behavior for certain variables.
In \S \ref{sec:Hydrodynamics} we present the equations for the
gauge invariant variables which have correct Newtonian limits, now, 
considering the general pressures in the background and perturbations.
We derive the general large scale solutions valid for general $K$ and $\Lambda$
in an ideal fluid medium (thus, valid for general equation of state
of the form $p = p(\mu)$, but with negligible entropic and anisotropic
pressures).
We also present a quantity which is conserved in the large scale
under general changes of the background equation of state for $K = 0$.
\S \ref{sec:Discussion} is a brief discussion.
We set $c \equiv 1$.

%%%%%%%%%%%%%%%%%%%%%%%%%%%%%%%%%%%%%%%%%%%%%%%%%%%%%%%%%%%%%%%%%
\section{Newtonian Cosmological Perturbations} 
                                    \label{sec:Newtonian}

The background evolution is governed by
\bea
   & & H^2 = {8 \pi G \over 3} \varrho - {K \over a^2} + {\Lambda \over 3}, 
       \quad \varrho \propto a^{-3},
   \label{Newt-BG}
\eea
where we allowed the general curvature (total energy) 
and the cosmological constant; $a(t)$ is a cosmic scale factor,
$H (t) \equiv \dot a / a$, and $\varrho (t)$ is the mass density. 
Perturbed parts of the mass conservation, the momentum conservation
and the Poisson's equations are (see eqs.[43,46] of H1): 
\bea
   & & \delta \dot \varrho + 3 H \delta \varrho = - {k \over a} \varrho
       \delta v, \quad
       \delta \dot v + H \delta v = {k \over a} \delta \Phi, \quad
       - {k^2 \over a^2} \delta \Phi = 4 \pi G \delta \varrho,
   \label{Newt-eqs} 
\eea
where $\delta \varrho ({\bf k}, t)$, $\delta v ({\bf k}, t)$ and
$\delta \Phi ({\bf k},t)$ are the Fourier mode of the perturbed mass density, 
velocity and gravitational potential, respectively.
Equation (\ref{Newt-eqs}) can be arranged into the closed form equations
for $\delta$ ($\equiv \delta \varrho / \varrho$), $\delta v$ and
$\delta \Phi$ as:
\bea
   & & \ddot \delta + 2 H \dot \delta - 4 \pi G \varrho \delta 
       = {1 \over a^2 H} \left[ a^2 H^2 \left( {\delta \over H} \right)^\cdot
       \right]^\cdot = 0,
   \label{Newt-delta-eq} \\
   & & \delta \ddot v + 3 H \delta \dot v + \left( \dot H + 2 H^2 
       - 4 \pi G \varrho \right) \delta v = 0,
   \label{Newt-v-eq} \\
   & & \delta \ddot \Phi + 4 H \delta \dot \Phi
       + \left( \dot H + 3 H^2 - 4 \pi G \varrho \right) \delta \Phi 
%       = \delta \ddot \Phi + 4 H \delta \dot \Phi
%       + \left( \Lambda - 2 {K \over a^2} \right) \delta \Phi 
       = {1 \over a^3 H} \left[ a^2 H^2 
       \left( {a \over H} \delta \Phi \right)^\cdot \right]^\cdot = 0.
   \label{Newt-Phi-eq} 
\eea
We note that equations (\ref{Newt-eqs})-(\ref{Newt-Phi-eq}) are
valid for general $K$ and $\Lambda$.
The general solutions for $\delta$, $\delta v$ and $\delta \Phi$
immediately follow as (see also Table 1 of H1):
\bea
   & & \delta ({\bf k},t) = (k^2 - 3K) \left[ 
       H C ({\bf k}) \int^t_0 {dt \over \dot a^2} 
       + {H \over 4 \pi G \varrho a^3} d ({\bf k}) \right], 
   \\
   & & \delta v ({\bf k}, t) = - {k^2 - 3K \over k^2} \left[ 
       {k \over aH} C ({\bf k})
       \left( 1 + a^2 H \dot H \int^t_0 {dt \over \dot a^2} \right)
       + {k \dot H \over 4 \pi G \varrho a^2} d ({\bf k}) \right],
   \\
   & & \delta \Phi ({\bf k}, t) = - {k^2 - 3K \over k^2}
       \left[ 4 \pi G \varrho a^2 H C ({\bf k}) \int^t_0 {dt \over \dot a^2}
       + {H \over a} d ({\bf k}) \right].
\eea
The $C$ and $d$ terms indicate the growing and decaying modes, respectively;
the coefficients are matched in accordance with the solutions with general
pressure in equations (\ref{varphi_chi-sol})-(\ref{v_chi-sol}).

%%%%%%%%%%%%%%%%%%%%%%%%%%%%%%%%%%%%%%%%%%%%%%%%%%%%%%%%%%%%%%%%%
\section{Relativistic Cosmological Perturbations} 
                                    \label{sec:Relativistic}

{}For the background we have (eq.[6] of Hwang 1993):
\bea
   H^2 = {8 \pi G \over 3} \mu - {K \over a^2} + {\Lambda \over 3}, \quad
       \dot \mu = - 3 H \left( \mu + p \right), \quad
       \dot H = - 4 \pi G ( \mu + p ) + {K \over a^2}.
   \label{BG-eqs} 
\eea
The third equation follows from the first two equations.
{}For $p = 0$ and replacing $\mu$ with $\varrho$ equation (\ref{BG-eqs}) 
reduces to equation (\ref{Newt-BG}).
The relativistic cosmological perturbation equations in a gauge ready form
were presented in Hwang (1991b).
We introduce familiar variables for the perturbed density and the perturbed 
velocity variables as
\bea
   \delta \equiv {\varepsilon \over \mu}, \quad
      v \equiv - {k \over a} {\Psi \over \mu + p}.
\eea
The gauge ready form of relativistic perturbation equations can be written as
(see eqs.[8-14] of Hwang 1993):
\bea
   & & \dot \varphi = H \alpha - {1\over 3} \kappa 
       + {1\over 3} {k^2 \over a^2} \chi,
   \label{eq1} \\
   & & - {k^2 - 3K \over a^2} \varphi + H \kappa = - 4 \pi G \mu \delta,
   \label{eq2} \\
   & & \kappa - {k^2 - 3K \over a^2} \chi = 12 \pi G \left( \mu + p \right)
       {a \over k} v,
   \label{eq3} \\
   & & \dot \chi + H \chi - \alpha - \varphi = 8 \pi G \sigma,
   \label{eq4} \\
   & & \dot \kappa + 2 H \kappa = \left( {k^2 \over a^2} - 3 \dot H \right) 
       \alpha + 4 \pi G \left( 1 + 3 c_s^2 \right) \mu \delta 
       + 12 \pi G e,
   \label{eq5} \\
   & & \dot \delta + 3 H \left( c_s^2 - {\rm w} \right) \delta 
       + 3 H { e \over \mu } = \left( 1 + {\rm w} \right)
       \left( \kappa - 3 H \alpha - {k \over a} v \right),
   \label{eq6} \\
   & & \dot v + \left( 1 - 3 c_s^2 \right) H v = {k \over a} \alpha
       + {k \over a \left( 1 + {\rm w} \right) } \left( c_s^2 \delta
       + {e \over \mu} - {2\over 3} {k^2 - 3K \over a^2} {\sigma \over \mu}
       \right).
   \label{eq7} 
\eea
According to their origins, equations (\ref{eq1})-(\ref{eq7}) can be 
called as the definition of $\kappa$, ADM energy constraint, 
momentum constraint, ADM propagation, Raychaudhuri, energy conservation,
and momentum conservation equations, respectively.
The perturbed metric variables $\varphi ({\bf k}, t)$, $\kappa ({\bf k},t)$,
$\chi ({\bf k}, t)$ and $\alpha ({\bf k}, t)$ are the perturbed part of
the three space curvature, expansion, shear and  lapse function, respectively.
The perturbed fluid variables $\delta ({\bf k}, t)$, $v ({\bf k}, t)$,
$e ({\bf k}, t)$ and $\sigma ({\bf k}, t)$ are the relative density
perturbation, (frame independent) velocity variable, entropic 
and anisotropic pressures, respectively.
The isotropic pressure is decomposed as
\bea
   & & \pi ({\bf k}, t) \equiv c_s^2 (t) \varepsilon ({\bf k}, t) 
       + e ({\bf k}, t), \quad
       c_s^2 \equiv {\dot p \over \dot \mu}, \quad
       {\rm w} (t) \equiv {p \over \mu}.
\eea
The perturbed variables used in equations (\ref{eq1})-(\ref{eq7})
are designed so that any one of the following conditions
can be used to fix the freedom based on the temporal gauge transformation:
$v \equiv 0$ (the comoving gauge), 
$\chi \equiv 0$ (the zero-shear gauge), 
$\kappa \equiv 0$ (the uniform-expansion gauge), 
$\alpha \equiv 0$ (the synchronous gauge), 
$\varphi \equiv 0$ (the uniform-curvature gauge),
and $\delta \equiv 0$ (the uniform-density gauge).
Each of these six gauge conditions, except for the synchronous gauge, 
fixes the temporal gauge transformation property completely.
Thus, each variable in these five gauge conditions 
is equivalent to a corresponding gauge invariant combination.
Due to the spatial homogeneity of the background, the effect from the 
spatial gauge transformation has been be trivially handled; $\chi$ is a
spatial gauge invariant combination, and the other metric and fluid variables 
are naturally spatially gauge invariant (Bardeen 1988).
The variables $e$ and $\sigma$ are gauge invariant.

We proposed a convenient way of writing the gauge invariant variables
(Hwang 1991b).
{}For example, we let
\bea
   & & \delta_v \equiv \delta + 3 (1 + {\rm w}) {aH \over k} v
       \equiv 3 (1 + {\rm w}) {a H \over k} v_\delta, 
   \nonumber \\
   & & \varphi_\chi \equiv \varphi - H \chi \equiv - H \chi_\varphi, \quad
       v_\chi \equiv v - {k \over a} \chi \equiv - {k \over a} \chi_v.
   \label{GI-combinations}
\eea
The variables $\delta_v$, $v_\chi$ and $\varphi_\chi$ are gauge invariant 
combination; $\delta_v$ becomes $\delta$ in the comoving gauge which takes
$v=0$ as the gauge condition, etc.
In this manner we can systematically construct the corresponding gauge 
invariant combination for any variable based on a gauge condition which 
fixes the temporal gauge transformation property completely.
{}For the gauge transformation properties, see \S 2.2 of Hwang (1991b).
A variable evaluated in different gauges can be considered as different 
variables, and they show different behaviors in general.
In this sense, except for the synchronous gauge, the variables in the 
rest of the five gauges can be considered as the gauge invariant variables.
(Thus, the variables with a subindex $\alpha$ are not gauge invariant, 
because those are equivalent to variables in the synchronous gauge.)
Although $\delta_v$ is a gauge invariant variable which becomes
$\delta$ in the comoving gauge, $\delta_v$ itself 
can be evaluated in any gauge with the same value.
The complete solutions for these six different gauge conditions are presented
in an ideal fluid case (Hwang 1993) and in a pressureless medium (H1).

In the following we {\it consider} a pressureless fluid with
$p = 0$ and $\pi = 0 = \sigma$.
Equations (\ref{eq1})-(\ref{eq7}) become:
\bea
   & & \dot \varphi = H \alpha - {1\over 3} \kappa 
       + {1\over 3} {k^2 \over a^2} \chi,
   \label{dust-1} \\
   & & - {k^2 - 3K \over a^2} \varphi + H \kappa = - 4 \pi G \mu \delta,
   \label{dust-2} \\
   & & \kappa - {k^2 - 3K \over a^2} \chi = 12 \pi G \mu {a \over k} v,
   \label{dust-3} \\
   & & \dot \chi + H \chi - \alpha - \varphi = 0,
   \label{dust-4} \\
   & & \dot \kappa + 2 H \kappa = \left( {k^2 \over a^2} - 3 \dot H \right) 
       \alpha + 4 \pi G \mu \delta,
   \label{dust-5} \\
   & & \dot \delta = \kappa - 3 H \alpha - {k \over a} v,
   \label{dust-6} \\
   & & \dot v + H v = {k \over a} \alpha.
   \label{dust-7}
\eea
In H1 and Hwang $\&$ Hyun (1994) we made arguments on the correspondences
between the Newtonian and the relativistic analyses.
In order to reinforce the Newtonian correspondences of certain
(gauge invariant) variables in certain gauges, in the following we
will present the closed form differential equations for 
$\delta$, $v$ and $\varphi$ in the six different gauge conditions.

%%%%%%%%%%%%%%%%%%%%%%%%%%%%%%%%%%%%%%%%%%%%%%%%%%%%%%%%%%%%%%%%%
\section{Equations in Six Fundamental Gauge Conditions} 
                                    \label{sec:Six-gauges}

In this section we consider a {\it pressureless medium}.
Thus, equations (\ref{dust-1})-(\ref{dust-7}) are the basic set of 
perturbation equations in a gauge ready form.

%%%%%%%%%%%%%%%%%%%%%%%%%%%%%%%%%%%%%%%%%%%%%%%%%%%%%%%%%%%%%%%%%
\subsection{Comoving Gauge} 
                                    \label{sec:CG}

As the gauge condition we set $v \equiv 0$.
Equivalently, we can set $v = 0$ and let every other variable as
the gauge invariant combinations with subindices $v$.
{}From equation (\ref{dust-7}) we have $\alpha_v = 0$.
Thus the comoving gauge is a case of the synchronous gauge; this is
true only in a pressureless situation.
{}From equations (\ref{dust-1})-(\ref{dust-7}) we can derive:
\bea
   & & \ddot \delta_v + 2 H \dot \delta_v - 4 \pi G \mu \delta_v = 0,
   \label{CG-delta-eq} \\
   & & \ddot \varphi_v + 3 H \dot \varphi_v - {K \over a^2} \varphi_v = 0.
   \label{CG-varphi-eq} 
\eea
Thus, equation (\ref{CG-delta-eq}) has the identical form as equation 
(\ref{Newt-delta-eq}).
We can show that the variables $a \kappa$ and $\chi/a$
satisfy the same equation for $\delta v$ in equation (\ref{Newt-v-eq});
see equations (\ref{SG-kappa-eq}) and (\ref{SG-chi-eq}).

{}For $K = 0$ equation (\ref{CG-varphi-eq}) leads to
two solutions which are $\varphi_v \propto {\rm constant}$ and 
$\int_0^t a^{-3} dt$. 
{}From equation (\ref{dust-7}) we have $\alpha_v = 0$.
Thus, for $K = 0$, from equations (\ref{dust-1}) and (\ref{dust-3})
we have $\dot \varphi_v = 0$.
This implies that, for $K = 0$, the second solution, 
$\varphi_v \propto \int_0^t a^{-3} dt$, should have the vanishing coefficient.
The general solution of equation (\ref{CG-varphi-eq}) will be
presented later in equation (\ref{varphi_v-sol3}).

%%%%%%%%%%%%%%%%%%%%%%%%%%%%%%%%%%%%%%%%%%%%%%%%%%%%%%%%%%%%%%%%%
\subsection{Zero-shear Gauge} 
                                    \label{sec:ZSG}

We let $\chi \equiv 0$, and substitute the other variables into the gauge
invariant combinations with subindices $\chi$.
We can derive:
\bea
   & & \ddot \delta_\chi + { 2 k^2 / a^2 - 36 \pi G \mu \left[ 1 + ( H^2 
       - 2 \dot H ) a^2 / \left( k^2 - 3 K \right) \right] \over
       k^2 / a^2 - 12 \pi G \mu \left[ 1 + 3 H^2 a^2/\left( k^2 - 3K \right) 
       \right] }
       H \left( \dot \delta_\chi 
       - 12 \pi G \mu {a^2 \over k^2 - 3 K} H \delta_\chi \right)
   \nonumber \\
   & & \qquad 
       - 4 \pi G \mu \left[ 1 - 3 ( H^2 - 2 \dot H )
       {a^2 \over k^2 - 3 K} \right] \delta_\chi = 0,
   \label{ZSG-delta-eq} \\
   & & \ddot v_\chi + 3 H \dot v_\chi 
       + \left( \dot H + 2 H^2 - 4 \pi G \mu \right) v_\chi = 0,
   \label{ZSG-v-eq} \\
   & & \ddot \varphi_\chi + 4 H \dot \varphi_\chi + \left( \dot H + 3 H^2 
       - 4 \pi G \mu \right) \varphi_\chi = 0.
   \label{ZSG-varphi-eq} 
\eea
Thus, equations (\ref{ZSG-v-eq}) and (\ref{ZSG-varphi-eq}) have identical forms 
as equations (\ref{Newt-v-eq}) and (\ref{Newt-Phi-eq}), respectively.
Only in the small scale limit the behavior of $\delta$ is the same 
as the Newtonian one.

%%%%%%%%%%%%%%%%%%%%%%%%%%%%%%%%%%%%%%%%%%%%%%%%%%%%%%%%%%%%%%%%%
\subsection{Uniform-expansion Gauge} 
                                    \label{sec:UEG}

We let $\kappa \equiv 0$, and substitute the remaining variables into
the gauge invariant combinations with subindices $\kappa$.
We can derive:
\bea
   & & \ddot \delta_\kappa 
       + \left( 2 H - {12 \pi G \mu H \over k^2/a^2 - 3 \dot H}
       \right) \dot \delta_\kappa - \left[ 4 \pi G \mu {k^2/a^2 + 6 H^2 \over
       k^2/a^2 - 3 \dot H} + \left( {12 \pi G \mu H \over
       k^2/a^2 - 3 \dot H} \right)^\cdot \right] \delta_\kappa = 0,
   \label{UEG-delta-eq} \\ 
   & & \ddot v_\kappa + \left[ 2 H - {12 \pi G \mu H \over k^2/a^2 - 3 \dot H}
       + {4 \pi G \mu \over k^2/a^2 - 3 \dot H}
       \left( {k^2/a^2 - 3 \dot H \over 4 \pi G \mu} \right)^\cdot
       \right] \dot v_\kappa
   \nonumber \\ 
   & & \quad 
       + \left[ \dot H + H^2 - 4 \pi G \mu {k^2/a^2 + 3 H^2 \over
       k^2/a^2 - 3 \dot H} + H {4 \pi G \mu \over k^2/a^2 - 3 \dot H}
       \left( { k^2/a^2 - 3 \dot H \over 4 \pi G \mu} \right)^\cdot
       \right] v_\kappa = 0,
   \label{UEG-v-eq} \\ 
   & & \ddot \varphi_\kappa 
       + \left( 4 H - {12 \pi G \mu H \over k^2/a^2 - 3 \dot H}
       \right) \dot \varphi_\kappa
   \nonumber \\ 
   & & \quad 
       + \left[ \dot H + 3 H^2 - 4 \pi G \mu {k^2/a^2 + 9 H^2 \over
       k^2/a^2 - 3 \dot H} - \left( {12 \pi G \mu H \over k^2/a^2
       - 3 \dot H} \right)^\cdot \right] \varphi_\kappa = 0.
   \label{UEG-varphi-eq}
\eea
In the small scale limit we can show that
equations (\ref{UEG-delta-eq})-(\ref{UEG-varphi-eq}) 
reduce to equations (\ref{Newt-delta-eq})-(\ref{Newt-Phi-eq}).
Thus, {\it in the small scale limit}, all three variables $\delta_\kappa$, 
$v_\kappa$ and $\varphi_\kappa$ correctly reproduce the Newtonian behavior.
However, outside or near horizon scale, the behaviors of all these
variables strongly deviate from the Newtonian ones.

In \S 84 of Peebles (1980) we find that in order to get the usual 
Newtonian equations a coordinate transformation is made so that
we have $\dot h \equiv 0$ in the new coordinate.
We can show that $\dot h = 2 \kappa$ in our notation.
Thus the new coordinate is in fact the uniform-expansion gauge.
\footnote{This was incorrectly pointed out below eq.[49] of H1;
in H1 it was mentioned that in \S 84 of Peebles (1980) the gauge 
transformations were made into the comoving gauge for $\delta$ and into the 
zero-shear gauge for $v$ and $\varphi$, respectively.}

%%%%%%%%%%%%%%%%%%%%%%%%%%%%%%%%%%%%%%%%%%%%%%%%%%%%%%%%%%%%%%%%%
\subsection{Synchronous Gauge} 
                                    \label{sec:SG}

We let $\alpha \equiv 0$.
This gauge condition does not fix the temporal gauge transformation property
completely.
Thus, although we can still indicate the variables in this gauge condition
using subindices $\alpha$ without ambiguity, these variables are not gauge 
invariant (see \S 3.2.1 of Hwang 1991b).
Equation (\ref{dust-7}) leads to
\bea
   v_\alpha = c_g {k \over a},
\eea
which is a pure gauge mode.
Thus, fixing $c_g \equiv 0$ exactly corresponds to the comoving gauge.
We can show that the following two equations are not affected by the
remaining gauge mode.
\bea
   & & \ddot \delta_\alpha + 2 H \dot \delta_\alpha 
       - 4 \pi G \mu \delta_\alpha = 0,
   \label{SG-delta-eq} \\
   & & \ddot \varphi_\alpha + 3 H \dot \varphi_\alpha 
       - {K \over a^2} \varphi_\alpha = 0.
   \label{SG-varphi-eq} 
\eea
Equation (\ref{SG-delta-eq}) is identical to equation (\ref{Newt-delta-eq}).
This is because the behavior of the gauge mode happens to coincide with 
the behavior of one of the physical mode for $\delta_\alpha$ and 
$\varphi_\alpha$.
However, for the variables $\kappa_\alpha$ and $\chi_\alpha$ the gauge mode 
contribution appears explicitly.
We can derive:
\bea
   & & (a \kappa_\alpha)^{\cdot \cdot} + 3 H (a \kappa_\alpha)^\cdot
       + \left( \dot H + 2 H^2 - 4 \pi G \mu \right) (a \kappa_\alpha) 
       = - 4 \pi G \mu k v_\alpha
       = - 4 \pi G \mu {k^2 \over a} c_g,
   \label{SG-kappa-eq} \\
   & & (\chi_\alpha/a)^{\cdot \cdot} + 3 H (\chi_\alpha/a)^\cdot
       + \left( \dot H + 2 H^2 - 4 \pi G \mu \right) (\chi_\alpha/a) 
       = - 4 \pi G \mu {1 \over k} v_\alpha.
%       = - 4 \pi G \mu {1 \over a} c_g.
   \label{SG-chi-eq}
\eea
In the comoving gauge the right hand sides of both equations vanish.

Thus, in a pressureless medium, variables
in the synchronous gauge behave the same as the ones in the comoving gauge, 
except for the additional gauge modes which appear in the synchronous gauge.
In a pressureless medium, we can simultaneously impose both the comoving gauge
and the synchronous gauge conditions.
However, this is possible only in a pressureless medium.
(see \S \ref{sec:Hydrodynamics}).

%%%%%%%%%%%%%%%%%%%%%%%%%%%%%%%%%%%%%%%%%%%%%%%%%%%%%%%%%%%%%%%%%
\subsection{Uniform-curvature Gauge} 
                                    \label{sec:UCG}

We let $\varphi \equiv 0$, and substitute the other variables into the gauge
invariant combinations with subindices $\varphi$.
We have
\bea
   & & \ddot \delta_\varphi + 2 H { (k^2 - 3K)/a^2 + 18 \pi G \mu \over 
       (k^2 - 3K)/a^2 + 12 \pi G \mu } \dot \delta_\varphi
       - { 4 \pi G \mu k^2/a^2 \over 
       (k^2 - 3K)/a^2 + 12 \pi G \mu } \delta_\varphi = 0,
   \label{UCG-delta-eq} \\
   & & \ddot v_\varphi + \left( 5 H + 2 {\dot H \over H} \right) \dot v_\varphi
       + \left( 3 \dot H + 4 H^2 \right) v_\varphi = 0. 
   \label{UCG-v-eq}
\eea
In the small scale limit equation (\ref{UCG-delta-eq}) reduces to
equation (\ref{Newt-delta-eq}).
In the uniform-curvature gauge, the perturbed potential is set equal
to zero by the gauge condition.
The uniform-curvature gauge condition has distinguished properties
in handling the scalar field or the dilaton field which appears in
some very general classes of the generalized gravity theories
(Hwang 1994b; Hwang $\&$ Noh 1996).

%%%%%%%%%%%%%%%%%%%%%%%%%%%%%%%%%%%%%%%%%%%%%%%%%%%%%%%%%%%%%%%%%
\subsection{Uniform-density Gauge} 
                                    \label{sec:UDG}

We let $\delta \equiv 0$, and substitute the other variables into the 
gauge invariant combinations with subindices $\delta$.
We have
\bea
   & & \ddot v_\delta + 2 \left( 2 H + {\dot H \over H} \right) \dot v_\delta
       + 3 \left( H^2 + \dot H \right) v_\delta = 0,
   \label{UDG-v-eq} \\
   & & \ddot \varphi_\delta + 2 H { (k^2 - 3 K)/a^2 + 18 \pi G \mu
       \over (k^2- 3K)/a^2 + 12 \pi G \mu} \dot \varphi_\delta
       - {4 \pi G \mu k^2/a^2
       \over (k^2- 3K)/a^2 + 12 \pi G \mu} \varphi_\delta = 0.
   \label{UDG-varphi-eq} 
\eea
These equations differ from equations (\ref{Newt-v-eq}) and (\ref{Newt-Phi-eq}).
Of course, we have no equation for $\delta$ which is set equal to zero 
by our choice of the gauge condition.

%%%%%%%%%%%%%%%%%%%%%%%%%%%%%%%%%%%%%%%%%%%%%%%%%%%%%%%%%%%%%%%%%
\section{Newtonian Correspondences} 
                                    \label{sec:Correspondences}

We found that equations for $\delta$ in the comoving gauge ($\delta_v$), and
for $v$ and $\varphi$ in the zero-shear gauge ($v_\chi$ and $\varphi_\chi$)
show the same forms as the corresponding Newtonian equations.
Using the gauge invariant combinations in equation (\ref{GI-combinations}),
equations (\ref{dust-2}), (\ref{dust-3}), (\ref{dust-4}), (\ref{dust-6})
and (\ref{dust-7}) can be combined to give:
\bea
   \dot \delta_v = - {k^2 - 3K \over a k} v_\chi, \quad
       \dot v_\chi + H v_\chi = - {k \over a} \varphi_\chi, \quad
       {k^2 - 3K \over a^2} \varphi_\chi = 4 \pi G \mu \delta_v.
   \label{GI-eqs}
\eea
Comparing equation (\ref{GI-eqs}) with equation (\ref{Newt-eqs}) 
we can identify the following correspondences:
\bea
   & & \delta_v \leftrightarrow \delta |_{\rm Newtonian}, \quad
       {k^2 - 3 K \over k^2} v_\chi 
       \leftrightarrow \delta v |_{\rm Newtonian}, \quad
       - {k^2 - 3 K \over k^2} \varphi_\chi \leftrightarrow 
       \delta \Phi |_{\rm Newtonian}.
   \label{Correspondences}
\eea
{}For the potential and density variables we can also
identify variables which behave similarly
\bea
   & & \delta_v = 3 H {a \over k} v_\delta, \quad
       v_\chi = - {k \over a} \chi_v = - {ak \over k^2 - 3 K} \kappa_v, \quad
       \varphi_\chi = -\alpha_\chi = - H \chi_\varphi,
\eea
where we used equations 
(\ref{GI-combinations}), (\ref{dust-3}), (\ref{dust-4}) and (\ref{GI-eqs}).

{}From a given solution we can derive all the rest of the solutions
even in other gauge conditions.
This can be done either by using the gauge invariant combination of variables 
or directly through gauge transformations.
General solutions in a pressureless medium are presented in H1.
{}From the complete solutions presented in Tables 1 and 2 of H1 we can
identify variables in certain gauges which correspond to the Newtonian ones.
These are summarized in Table 1.

\centerline{ [[{\bf TABLE 1.}]] }

\noindent
{}From Table 1 we notice that in the small scale limit, except for the  
uniform-density gauge where $\delta \equiv 0$, 
$\delta$ in all gauge conditions behaves in the same way.
Thus, as in equation (\ref{Correspondences}), it may be natural to identify 
$\delta$ in the comoving gauge with the corresponding Newtonian one.

Notice that, although we have horizons in the relativistic analysis
the equations for $\delta_v$, $v_\chi$ and $\varphi_\chi$
keep the same form as the corresponding Newtonian equations
{\it in the general scale}.
Considering this as an additional point we regard
$\delta_v$, $v_\chi$ and $\varphi_\chi$ as most closely corresponding
ones to the Newtonian variables.

%%%%%%%%%%%%%%%%%%%%%%%%%%%%%%%%%%%%%%%%%%%%%%%%%%%%%%%%%%%%%%%%%
\section{Relativistic Cosmological Hydrodynamics}
                                \label{sec:Hydrodynamics}

In the previous sections we have shown that the gauge invariant
combinations $\delta_v$, $v_\chi$ and $\varphi_\chi$ behave most similarly
to the Newtonian $\delta \equiv \delta \varrho/\varrho$, 
$\delta v$ and $\delta \Phi$.
The equations remain the same in a general scale which includes the
superhorizon scales in the relativistic situation.
In this section, we will present the equations for $\delta_v$, $v_\chi$ and
$\varphi_\chi$ including the effects of the general pressure terms.
Equations (\ref{eq1})-(\ref{eq7}) are the basic set of 
perturbation equations in a gauge ready form.

{}From equations (\ref{eq3}), (\ref{eq6}) and (\ref{eq7}) we have
\bea
   \dot \delta_v - 3 H {\rm w} \delta_v
      = - \left( 1 + {\rm w} \right) {k \over a} {k^2 - 3 K \over k^2} v_\chi
      - 2 H {k^2 - 3K \over a^2} {\sigma \over \mu}.
   \label{P-1}
\eea
{}From equations (\ref{eq4}) and (\ref{eq7}) we have
\bea
   \dot v_\chi + H v_\chi = - {k \over a} \varphi_\chi
      + {k \over a \left( 1 + {\rm w} \right)} 
      \left[ c_s^2 \delta_v + {e \over \mu}
      - 8 \pi G \left( 1 + {\rm w} \right) \sigma
      - {2 \over 3} {k^2 - 3 K \over a^2} {\sigma \over \mu} \right].
   \label{P-2}
\eea
{}From equations (\ref{eq2}) and (\ref{eq3}) we have
\bea
   {k^2 - 3 K \over a^2} \varphi_\chi = 4 \pi G \mu \delta_v.
   \label{P-3}
\eea
{}From equations (\ref{eq1}), (\ref{eq3}) and (\ref{eq4}) we have
\bea
   \dot \varphi_\chi + H \varphi_\chi
      = - 4 \pi G \left( \mu + p \right) {a \over k} v_\chi - 8 \pi G H \sigma.
   \label{P-4}
\eea
Considering the correspondences in equation (\ref{Correspondences})
we can immediately see that equations (\ref{P-1})-(\ref{P-4}) have
the correct Newtonian limit expressed in equations (\ref{Newt-eqs}).

Combining equations (\ref{P-1})-(\ref{P-4}) we can derive the closed
form expressions for the $\delta_v$, $v_\chi$ and $\varphi_\chi$
which are the relativistic counterpart of equations 
(\ref{Newt-delta-eq})-(\ref{Newt-Phi-eq}).
We have:
\bea
   & & \ddot \delta_v + \left( 2 + 3 c_s^2 - 6 {\rm w} \right) H \dot \delta_v
       + \Bigg[ c_s^2 {k^2 \over a^2} 
       - 4 \pi G \mu \left( 1 - 6 c_s^2 + 6 {\rm w} - 3 {\rm w}^2 \right)
%   \nonumber \\
%   & & \qquad \qquad \qquad
       + 12 \left( {\rm w} - c_s^2 \right) {K \over a^2}
       + \left( 3 c_s^2 - 5 {\rm w} \right) \Lambda \Bigg] \delta_v
   \nonumber \\
   & & \qquad
       = {1 + {\rm w} \over a^2 H} \left[ {H^2 \over a (\mu + p)} 
       \left( {a^3 \mu \over H} \delta_v \right)^\cdot \right]^\cdot
       + c_s^2 {k^2 \over a^2} \delta_v
   \nonumber \\
   & & \qquad
       = - {k^2 - 3 K \over a^2} { 1 \over \mu }
       \left\{ e + 2 H \dot \sigma
       + 2 \left[ - {1 \over 3} {k^2 \over a^2} + 2 \dot H
       + 3 \left( 1 + c_s^2 \right) H^2 \right] \sigma \right\},
   \label{delta-eq}\\
   & & \ddot \varphi_\chi + \left( 4 + 3 c_s^2 \right) H \dot \varphi_\chi
       + \left[ c_s^2 {k^2 \over a^2} 
       + 8 \pi G \mu \left( c_s^2 - {\rm w} \right)
       - 2 \left( 1 + 3 c_s^2 \right) {K \over a^2}
       + \left( 1 + c_s^2 \right) \Lambda \right] \varphi_\chi
   \nonumber \\
   & & \qquad
       = {\mu + p \over H} \left[ {H^2 \over a (\mu + p)} 
       \left( {a \over H} \varphi_\chi \right)^\cdot \right]^\cdot
       + c_s^2 {k^2 \over a^2} \varphi_\chi
   \nonumber \\
   & & \qquad
       = - 4 \pi G \left\{ e + 2 H \dot \sigma
       + 2 \left[ - {1 \over 3} {k^2 \over a^2} + 2 \dot H
       + 3 \left( 1 + c_s^2 \right) H^2 \right] \sigma \right\}.
   \label{varphi-eq}
\eea
It may be an interesting exercise to show that the above equations
are indeed valid for general $K$ and $\Lambda$, and
for the general equation of state $p = p(\mu)$;
use $\dot {\rm w} = - 3 H (1 + {\rm w}) ( c_s^2 - {\rm w} )$.

If we ignore the entropic and anisotropic pressures ($e = 0 = \sigma$) 
on scales larger than Jeans scale
equation (\ref{varphi-eq}) immediately leads to a general integral form 
solution as (see \S 3.2.3 of Hwang 1991b, and \S V of Hwang $\&$ Vishniac 1990)
\bea
   \varphi_\chi ({\bf k}, t)
   &=& 4 \pi G C ({\bf k}) {H \over a} \int_0^t {a (\mu + p) \over H^2} dt
       + {H \over a} d ({\bf k})
   \nonumber \\
   &=& C ({\bf k}) \left[ 1 - {H \over a} \int_0^t
       a \left( 1 - {K \over \dot a^2} \right) dt \right]
       + {H \over a} d ({\bf k}).
   \label{varphi_chi-sol}
\eea
Solutions for $\delta_v$ and $v_\chi$ follow from equations
(\ref{P-3}) and (\ref{P-4}), respectively, as
\footnote{We correct a typographical error in equation (129) of 
Hwang $\&$ Vishniac: $c_0/a$ should be replaced by $c_0$.}
\bea
   & & \delta_v ({\bf k}, t) = {k^2 - 3 K \over 4 \pi G \mu a^2} 
       \varphi_\chi ({\bf k}, t), 
   \label{delta_v-sol} \\
   & & v_\chi ({\bf k}, t)= - {k \over 4 \pi G ( \mu + p) a^2}
       \left\{ C ({\bf k}) \left[ {K \over \dot a} 
       - \dot H \int_0^t a \left( 1 - {K \over \dot a^2} \right) dt \right]
       + \dot H d ({\bf k}) \right\}.
   \label{v_chi-sol}
\eea

We stress that these large scale asymptotic solutions are
valid for general $K$ and $\Lambda$, and for general $p = p(\mu)$.
{}For $K = 0 = \Lambda$ and ${\rm w} = {\rm constant}$ we have
$a \propto t^{2/[3(1+{\rm w})]}$ and equations 
(\ref{varphi_chi-sol})-(\ref{v_chi-sol}) become:
\bea
   & & \varphi_\chi ({\bf k}, t) 
       = {3 (1 + {\rm w}) \over 5 + 3 {\rm w}} C ({\bf k})
       + {2 \over 3 (1 + {\rm w})} {1 \over at} d ({\bf k})
       \quad \propto \quad {\rm constant}, 
       \quad t^{- {5 + 3 {\rm w} \over 3 ( 1 + {\rm w} )} },
   \nonumber \\
   & & \delta_v ({\bf k}, t) 
       \quad \propto \quad t^{2(1 + 3 {\rm w}) \over 3 (1 + {\rm w})}, 
       \quad t^{- {1 - {\rm w} \over 1 + {\rm w} } },
   \nonumber \\
   & & v_\chi ({\bf k}, t)
       \quad \propto \quad t^{1 + 3 {\rm w} \over 3 (1 + {\rm w})}, 
       \quad t^{- {4 \over 3(1 + {\rm w}) } }.
   \label{Sol-w=const}
\eea
We note that $C({\bf k})$ and $d({\bf k})$ are integration constants which are
independent of the temporal evolution of the background equation of state,
i.e., constants for general $p = p(\mu)$.
      
The equations in this section and equations (\ref{BG-eqs})-(\ref{eq7})
include general pressures which may account for the nonequilibrium or
dissipative effects in hydrodynamic flows in the cosmological context
(with general $K$ and $\Lambda$).
The equations are expressed in general forms so that
they can include the case of the scalar field and classes
of generalized gravity theories.
Application of the gauge ready formalism to the minimally coupled scalar
field was made in Hwang (1994b), and to the generalized gravity theories
in Hwang $\&$ Noh (1996).

%%%%%%%%%%%%%%%%%%%%%%%%%%%%%%%%%%%%%%%%%%%%%%%%%%%%%%%%%%%%%%%%%
\subsection{A Conserved Quantity}
                                \label{sec:Conservation}

It is known that the curvature fluctuation in the comoving gauge, $\varphi_v$, 
is conserved in the large scale limit independently
of the changes in the background equation of state (Hwang $\&$ Hyun 1994).
{}From equation (\ref{GI-combinations}) we have
\bea
   \varphi_v = \varphi_\chi - {aH \over k} v_\chi.
\eea
Thus, from equations (\ref{varphi_chi-sol}) and (\ref{v_chi-sol}) we can derive
\bea
   \varphi_v ({\bf k}, t) 
   &=& C ({\bf k}) \left\{ 1 + {K \over a^2}
       {1 \over 4 \pi G (\mu + p)} \left[ 1 
       - {H \over a} \int_0^t a \left( 1 - {K \over \dot a^2} \right) dt 
       \right] \right\}
       + {K \over a^2} { H/a \over 4 \pi G ( \mu + p)} d ({\bf k})
   \nonumber \\
   &=& C ({\bf k}) \left[ 1 + {K \over a^2}
       {H/a \over \mu + p} \int_0^t {a (\mu + p) \over H^2} dt \right]
       + {K \over a^2} { H/a \over 4 \pi G ( \mu + p)} d ({\bf k}).
   \label{varphi_v-sol}
\eea
{}For $K = 0$ (but for general $\Lambda$) we have
\bea
   \varphi_v ({\bf k}, t) = C ({\bf k}),
   \label{varphi_v-sol2}
\eea
with the vanishing decaying mode;
the disappearance of the decaying mode in equation (\ref{varphi_v-sol2})
implies that the dominating decaying modes in equations (\ref{varphi_chi-sol}) 
and (\ref{v_chi-sol}) cancel out for $K = 0$.
Thus, for $K = 0$, $\varphi_v$ is conserved for generally varying
background equation of state, i.e., for general $p = p(\mu)$.
This conservation property of the curvature variable in a certain gauge
remains to be true for models which are based on a minimally coupled scalar 
field or even on classes of generalized gravity theories; 
in the generalized gravity the uniform-field gauge is more suitable
for handling the conservation property, and the uniform-field gauge
coincides with the comoving gauge in the minimally coupled scalar field
(Hwang 1994b and Hwang $\&$ Noh 1996).
{}For a pressureless medium equation (\ref{varphi_v-sol}) reduces to
\bea
   \varphi_v ({\bf k},t) = C({\bf k}) 
      \left( 1 + K H \int_0^t {dt \over \dot a^2} \right)
      + K H { 1 \over 4 \pi G \mu a^3} d ({\bf k}).
   \label{varphi_v-sol3}
\eea

%%%%%%%%%%%%%%%%%%%%%%%%%%%%%%%%%%%%%%%%%%%%%%%%%%%%%%%%%%%%%%%%%
\section{Discussion}
                                \label{sec:Discussion}

We would like to make comments on related works in the books by Weinberg (1972) 
and Peebles (1993).
Equation (15.10.57) in Weinberg (1972) and equation (10.118) in Peebles (1993)
are in error.
The correction in the case of Weinberg's was made in Hwang (1991a); 
in a medium with nonvanishing pressure the equation for the density fluctuation
in the synchronous gauge becomes a third order because of the presence of 
a gauge mode in addition to the physical growing and decaying modes.
The truncated second order equations in Weinberg and Peebles will pick up 
a gauge mode instead of the physical decaying mode in the synchronous gauge.
The error in Peebles is based on imposing the synchronous gauge and the
comoving gauge simultaneously.
In a medium with nonvanishing pressure one cannot impose the two gauge 
conditions simultaneously even in the large scale limit.  
In the Appendix we elaborate our point.

In this paper we have tried to identify the variables in the relativistic
perturbation analysis which reproduce the correct Newtonian behavior
in the pressureless limit.
In \S \ref{sec:Six-gauges} we have shown that $\delta$, $\delta v$ 
and $\varphi$ in the uniform-expansion gauge reproduce the Newtonian 
behaviors in the small scale limit 
(i.e., on scales smaller than the visual horizon).
However, the variables change their behaviors near and outside horizon scale.
Meanwhile, $\delta$ in the comoving gauge and $v$ and $\varphi$ in the 
zero-shear gauge show the same behavior as the corresponding Newtonian 
variables {\it in a general scale}.
In the small scale limit the density peturbation in most of the gauge 
conditions correctly reproduces the Newtonian behavior.
In fact, these results were already presented in H1. 
Comparing with H1, in the present work we tried to reinforce the 
correspondence by showing explicitly the second order differential equations 
which are satisfied by the variables in several gauge conditions.
Various general and asymptotic solutions for every variable
in the pressureless medium are presented in the Tables of H1.

%%%%%%%%%%%%%%%%%%%%%%%%%%%%%%%%%%%%%%%%%%%%%%%%%%%%%%%%%%%%%%%%%
JH wishes to thank Prof. R. Brandenberger for interesting correspondences.
This work was supported by the KOSEF, Grants No. 95-0702-04-3 and
No. 961-0203-013-1, and through the SRC program of SNU-CTP.

%%%%%%%%%%%%%%%%%%%%%%%%%%%%%%%%%%%%%%%%%%%%%%%%%%%%%%%%%%%%%%%%%
\section*{Appendix: Analysis in the synchronous gauge.}

In the following we correct a minor confusion in the literature
concerning perturbation analyses in the synchronous gauge.
The argument is based on Hwang (1991a; H2 hereafter).
We consider a situation with $K = 0 = \Lambda$, $e = 0 = \sigma$ and 
${\rm w} = {\rm constant}$ (thus $c_s^2 = {\rm w}$).

The equation for the density perturbation in the {\it comoving gauge}
is given in equation (\ref{delta-eq}).
In our case we have
$$
   \ddot \delta_v + ( 2- 3 {\rm w}) H \dot \delta_v
      + \left[ c_s^2 {k^2 \over a^2}
      - 4 \pi G \mu (1 - {\rm w}) ( 1 + 3 {\rm w} ) \right] \delta_v = 0.
   \eqno{(A1)}
$$
The solution in the large scale limit is presented in 
equation (\ref{Sol-w=const}).

In the {\it synchronous gauge}, from equations (\ref{eq5})-(\ref{eq7}) we have
$$
   \ddot \delta_\alpha + 2 H \dot \delta_\alpha
       + \left[ c_s^2 {k^2 \over a^2}
       - 4 \pi G \mu ( 1+ {\rm w} ) ( 1 + 3 {\rm w} ) \right] \delta_\alpha
       + 3 {\rm w} ( 1 + {\rm w}) {k \over a} H v_\alpha = 0,
   \eqno{(A2)}
$$
$$
   \dot v_\alpha + ( 1- 3 {\rm w}) H v_\alpha
       - {k \over a} {{\rm w} \over 1 + {\rm w}} \delta_\alpha = 0.
   \eqno{(A3)}
$$
{}From these two equations we can derive a third order differential equation
for $\delta_\alpha$ as
$$
   \delta^{\cdot\cdot\cdot}_\alpha
       + {11 - 3 {\rm w} \over 2} H \ddot \delta_\alpha
       + \left( {5 - 24 {\rm w} - 9 {\rm w}^2 \over 2} H^2
       + c_s^2 {k^2 \over a^2} \right) \dot \delta_\alpha
$$
$$
       + \left[ {3 \over 4} \left( 1 + {\rm w} \right)
       \left( 1 + 3 {\rm w} \right) \left( -1 + 9 {\rm w} \right) H^3
       + {3 \over 2} \left( 1 + {\rm w} \right) H c_s^2 {k^2 \over a^2}
       \right] \delta_\alpha = 0.
   \eqno{(A4)} 
$$
In the large scale limit the solutions are
$$
   \delta_\alpha \quad \propto \quad
       t^{{ 2(1+ 3{\rm w}) \over 3(1 + {\rm w})}}, \quad
       t^{{9{\rm w} - 1 \over 3( 1+ {\rm w})}}, \quad
       t^{-1}.
   \eqno{(A5)}
$$
In the Appendix B of H2 we made an argument that 
the third solution with $\delta_\alpha \propto t^{-1}$
is nothing but a gauge mode for a medium with ${\rm w} \neq 0$.
({}For a pressureless case the physical decaying mode also behaves as $t^{-1}$
and the second solution in eq.[A5] is invalid; see \S 4 of H2).
(As mentioned before, the combination $\delta_\alpha$ is not gauge
invariant. 
We have $\delta_\alpha \equiv \delta + 3 H ( 1 + {\rm w}) \int^t \alpha dt$
and the lower bound of the integration gives rise to the gauge mode, thus
behaving as proportional to $t^{-1}$; see eq.[44] of Hwang 1991b).
In Eq. (B5) of of H2 we derived the relation between solutions
in the two gauges explicitly; the growing modes are the same in both gauges
whereas the decaying modes differ by a factor $(k / aH)^2$.

Now, we would like to point out that even in the large scale limit
one cannot ignore the last term in equation (A2).
{\it If we ignore the last term} in equation (A2) we recover 
equation (15.10.57) in Weinberg (1972) and equation (10.118) in Peebles (1993)
which is
$$
   \ddot \delta_\alpha + 2 H \dot \delta_\alpha
       + \left[ c_s^2 {k^2 \over a^2} 
       - 4 \pi G \mu ( 1 + {\rm w} ) ( 1 + 3 {\rm w} ) \right] \delta_\alpha 
       = 0.
   \eqno{(A6)}
$$
In the large scale we have solutions
$$
   \delta_\alpha \quad \propto \quad
       t^{{ 2(1+ 3{\rm w}) \over 3(1 + {\rm w})}}, \quad
       t^{-1}.
   \eqno{(A7)}
$$
By ignoring the last term in equation (A2), in the large scale
limit we happen to recover the fictitious gauge mode under the price of losing 
the physical decaying mode.
Thus, in the large scale limit one cannot ignore the last term in equation (A2)
in a medium with general pressure; this is apparent in equation (A4).
Also, one cannot impose both the synchronous gauge condition and the 
comoving gauge condition simultaneously.  
If we simultaneously impose such two conditions, 
thus setting $v \equiv 0 \equiv \alpha$, from equation (\ref{eq7}) we have
$$
   0 = {k \over a \left( 1 + {\rm w} \right) } \left( c_s^2 \delta
       + {e \over \mu} - {2\over 3} {k^2 - 3K \over a^2} {\sigma \over \mu}
       \right).
   \eqno{(A8)}
$$
Thus, for $e = 0 = \sigma$ and medium with nonvanishing pressure 
we have $\delta = 0$ which is a meaningless system; 
this argument remains valid even in the large scale limit.

%%%%%%%%%%%%%%%%%%%%%%%%%%%%%%%%%%%%%%%%%%%%%%%%%%%%%%%%%%%%%%%%%
\section*{References}

\n Bardeen, J. M. 1980, Phys. Rev. D, 22, 1882

\n -----------. 1988, in Particle Physics and Cosmology, ed. A. Zee (London: 
   Gordon $\&$ Breach), 1p

\n Bonner, W. B. 1957, MNRAS, 107, 104

\n Hwang, J. 1991a, Gen. Rel. Grav., 23, 235 (H2)

\n -----------. 1991b, ApJ, 375, 443

\n -----------. 1993, ApJ, 415, 486

\n -----------. 1994a, ApJ, 427, 533 (H1)

\n -----------. 1994b, ApJ, 427, 542

\n Hwang, J., $\&$ Hyun, J. J. 1994, ApJ, 420, 512 

\n Hwang, J., $\&$ Noh, H. 1996, Phy. Rev. D, 54, 1460 

\n Hwang, J., $\&$ Vishniac, E. T. 1990, ApJ, 353, 1

\n Kodama, H., $\&$ Sasaki, M. 1984, Prog. Theor. Phys. Suppl., 78, 1

\n Lifshitz, E. M. 1946, J. Phys. (USSR), 10, 116

\n Nariai, H. 1969, Prog. Theor. Phys., 41, 686

\n Peebles, P. J. E. 1980, The Large-Scale Structure of the Universe (Princeton:
   Princeton Univ. Press)

\n -----------. 1993, Principles of Physical Cosmology (Princeton:
   Princeton Univ. Press)

\n Sakai, K. 1969, Prog. Theor. Phys., 41, 1461

\n Weinberg, S. 1972, Gravitation and Cosmology (Wiley: New York)

\vskip 2cm
%%%%%%%%%%%%%%%%%%%%%%%%%%%%%%%%%%%%%%%%%%%%%%%%%%%%%%%%%%%%%%%%%

\noindent
{\bf Table 1.  Newtonian correspondences:}
{}For the synchronous gauge we ignore the gauge mode. 
Thus the synchronous gauge is equivalent to the comoving gauge.
Dots (\dots) indicate that the behavior differs from the Newtonian one.
The small scale implies the scale smaller than the visual horizon.
Explicit forms of exact and asymptotic solutions for every variable are 
presented in H1.

\noindent
===============================================
\begin{tabbing}
Gauge \hskip 4.7cm\= Variable \hskip 1.3cm \= General Scale \hskip 1cm\= Small Scale\\
--------------------------------------------------------------------------------------------------------------  \\
Comoving gauge          \> $\delta_v$          \> Newtonian \> Newtonian  \\
Zero-shear gauge        \> $\delta_\chi$       \> \dots     \> Newtonian  \\
Uniform-expansion gauge \> $\delta_\kappa$     \> \dots     \> Newtonian  \\
Synchronous gauge       \> $\delta_\alpha$     \> Newtonian \> Newtonian  \\
Uniform-curvature gauge \> $\delta_\varphi$    \> \dots     \> Newtonian  \\ 
Uniform-density gauge   \> $\delta \equiv 0$   \> 0         \> 0          \\
--------------------------------------------------------------------------------------------------------------  \\
Comoving gauge          \> $v \equiv 0$        \> 0         \> 0          \\
Zero-shear gauge        \> $v_\chi$            \> Newtonian \> Newtonian  \\
Uniform-expansion gauge \> $v_\kappa$          \> \dots     \> Newtonian  \\
Synchronous gauge       \> $v_\alpha$          \> 0         \> 0          \\
Uniform-curvature gauge \> $v_\varphi$         \> \dots     \> \dots      \\ 
Uniform-density gauge   \> $v_\delta$          \> \dots     \> \dots      \\
--------------------------------------------------------------------------------------------------------------  \\
Comoving gauge          \> $\varphi_v$         \> \dots     \> \dots      \\
Zero-shear gauge        \> $\varphi_\chi$      \> Newtonian \> Newtonian  \\
Uniform-expansion gauge \> $\varphi_\kappa$    \> \dots     \> Newtonian  \\
Synchronous gauge       \> $\varphi_\alpha$    \> \dots     \> \dots      \\
Uniform-curvature gauge \> $\varphi \equiv 0$  \> 0         \> 0          \\ 
Uniform-density gauge   \> $\varphi_\delta$    \> \dots     \> \dots      \\
--------------------------------------------------------------------------------------------------------------  \\
\end{tabbing}

%%%%%%%%%%%%%%%%%%%%%%%%%%%%%%%%%%%%%%%%%%%%%%%%%%%%%%%%%%%%%%%%%%%
\end{document}